\documentstyle[aps,floats]{revtex}
\input{psfig.tex}
\begin{document}
\draft

\title{Seeing Sound Waves in the Early Universe}

\author{Arthur Kosowsky
}
\address{
Department of Physics and Astronomy, Rutgers University,
136 Frelinghuysen Road, Piscataway, New Jersey~~08854-8019
}
\date{October, 1998}
\maketitle

\begin{abstract}
Temperature and polarization power spectra of the cosmic
microwave background can provide essentially incontrovertible
evidence for coherent acoustic oscillations in the early universe.
A simple model calculation demonstrates explicitly
how polarization couples to velocities at the surface of last scatter
and is nearly independent of gravitational or density perturbations.
For coherent acoustic oscillations, 
peaks in the temperature and polarization power spectra
are precisely interleaved. If
observed, such a signal would provide strong support for initial
density perturbations on scales larger than the horizon, and thus
for inflation.
\end{abstract}


A hallmark of inflation-type cosmological models is density
perturbations on all scales, including those larger than the
causal horizon. Prior to recombination, a given Fourier
mode density perturbation begins oscillating as an acoustic
wave once the horizon overtakes its wavelength. Since all modes
with a given wavelength begin evolving simultaneously, the
resulting acoustic oscillations are phase-coherent, leading to
the familiar acoustic peaks (often termed, misleadingly,
``Doppler peaks'') in the temperature power spectrum of the
cosmic microwave background. These peaks give sufficient
structure to the power spectrum to enable precision
determination of cosmological parameters in this simple and
general class of models \cite{parameters}.

While a measured temperature power spectrum which matches a
particular inflation-type model would be quite strong evidence
in favor of the model, it would still constitute only indirect
evidence for acoustic oscillations, requiring a model fit. Even
more importantly, if the observed power spectrum has apparent 
acoustic peaks but is not well-fit by any simple model, the nature
and source of these temperature power spectrum features would be
open to question. This Letter points out that polarization of the 
microwave background, when combined with temperature information,
provides an intuitively clear and compelling physical signature
of coherent acoustic oscillations in the early universe.

Figure 1 displays the temperature and polarization power spectra
and the cross-correlation between temperature and polarization for
a typical inflationary model \cite{cmbfast}. The
multipole moment $\ell$ corresponds to an angle on the sky
$\theta\simeq\pi/\ell$ and to a flat-universe comoving length scale
$L\simeq 20000/\ell$ Mpc at the surface of last scattering which
is imaged by the microwave background. The peak structure in the
spectra result from coherent acoustic oscillations. 
Microwave background polarization couples
almost exclusively to velocity perturbations and not to density
perturbations, while the peaks in the temperature power spectrum
arise almost completely from density perturbations. The main point of
this paper is that for coherent acoustic oscillations,
the alternating temperature and polarization peaks are precisely
interleaved. The acoustic modes have the same initial phase but
different oscillation frequencies depending on the wavelength
of the mode. Thus at the surface of last scattering, the phase of
acoustic oscillations varies smoothly with wavelength, resulting
in alternating temperature (density) and polarization (velocity) peaks
as a function of $\ell$. As an additional check, the extrema in
the cross correlation must fall between the polarization and
temperature peaks. The remainder of this paper is devoted to
a quantitative derivation of this signature.

\begin{figure}[htbp] 
\centerline{\psfig{file=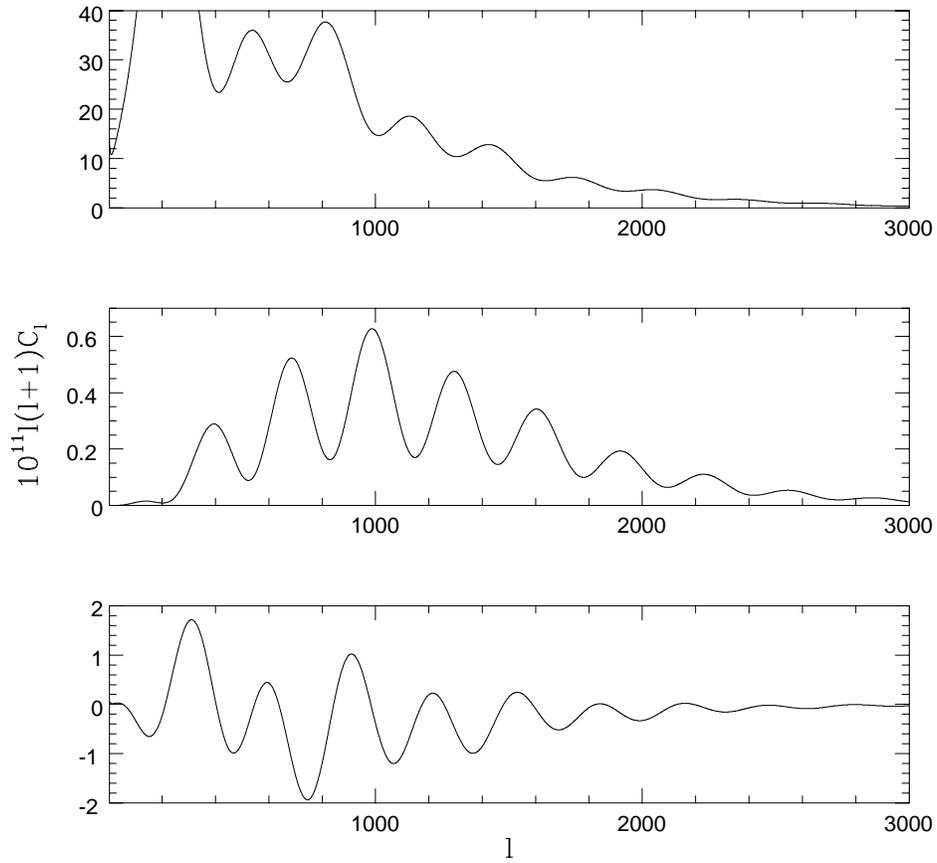,width=5in}}
\caption{The power spectra for temperature fluctuations (top),
polarization (center) and temperature-polarization cross-correlation
(bottom) for a typical inflationary model. The oscillations remain
in phase up to $l=3000$.
}
\end{figure}

Three basic effects must be understood: (1) Polarization is
generated through Thomson scattering, and only the quadrupole
moment of an incoming 
unpolarized radiation field can be scattered into a
polarized outgoing field; (2) The radiation field just before
decoupling has only monopole and dipole angular dependence, and
a non-zero quadrupole is generated through free-streaming of the
radiation; (3) The specific form of the acoustic modes fixes the
spatial dependence of the initial monopole and dipole radiation
fields. Combining these three ingredients results in the
acoustic signature in question. Each of these three points will
be considered in turn.

Polarization in the microwave background is generated through
polarization-dependent Thomson scattering.
Consider Thomson scattering of an incoming monochromatic
electromagnetic plane wave by an electron \cite{annals}. 
The total scattering cross-section
is given by
\begin{equation}
{d\sigma\over d\Omega} = {3\sigma_T\over 8\pi} 
\left|{\hat\varepsilon}'\cdot{\hat\varepsilon}\right|^2
\label{thomson}
\end{equation}
where $\sigma_T$ is the total Thomson cross section and
the vectors $\hat\varepsilon$ and ${\hat\varepsilon}'$ are unit
vectors in the planes perpendicular to the propagation directions
which are aligned with the outgoing and incoming linear polarization,
respectively. Unpolarized radiation may be considered as the 
superposition of two plane waves with equal amplitudes and phases but
orthogonal polarizations. A straightforward calculation gives the 
net polarization produced by the scattering of an unpolarized
radiation field of intensity $I'(\theta,\phi)$ incident on a
small volume containing an electron as
\begin{equation}
Q({\bf\hat z})-iU({\bf\hat z}) 
= {3\sigma_T\over 16\pi\sigma_B}\int d\Omega 
\sin^2\theta e^{2i\phi} I'(\Omega),
\label{stokesint}
\end{equation}
where $\sigma_B$ is the cross-sectional area of the small
test volume, the $z$-axis is taken as the scattering direction,
and $Q$ and $U$ are the Stokes parameters describing linear
polarization.
Expanding the incident radiation field in spherical harmonics,
\begin{equation}
I'(\theta,\phi) \equiv \sum_{lm}a_{lm}Y_{lm}(\theta,\phi),
\end{equation}
gives the outgoing Stokes parameters in terms of the multipole
coefficients as
\begin{equation}
Q({\bf\hat z})-iU({\bf\hat z}) 
={3\sigma_T\over 4\pi\sigma_B} \sqrt{2\pi\over 15}\,a_{22}.
\label{a22}
\end{equation}
Thus polarization is generated along the outgoing $z$-axis provided that
the $a_{22}$ quadrupole moment of the incoming radiation is non-zero. To
determine the outgoing polarization in an
arbitrary direction $\bf\hat n$ making an 
angle $\beta$
with the $z$-axis, the same physical incoming field must be multipole expanded
in a coordinate system rotated through the Euler angle $\beta$. 
If the incoming radiation field is independent of
$\phi$, as it will be for individual Fourier components of a density
perturbation, then
\begin{equation}
Q({\bf\hat n})-iU({\bf\hat n}) 
={3\sigma_T\over 16\pi\sigma_B} \sqrt{4\pi\over 5}\,a_{20}\sin^2\beta
\label{a20}
\end{equation}
which can be derived using explicit expressions
for the $l=2$ components of the rotation matrix.
In other words, an azimuthally-symmetric
radiation field will generate a polarized scattered field if
it has a non-zero $a_{20}$ multipole component, and the magnitude
of the scattered polarization will be proportional to $\sin^2\beta$.
Since the incoming field is real, $a_{20}$ will be real, $U=0$,
and the polarization orientation will be along longitudes of
the coordinate system \cite{primer}. 

Moving on to the second part of the calculation, where does an
incident quadrupole radiation field come from? Before decoupling,
the radiation is tightly coupled to the baryons in the universe by
rapid Thomson scattering and the radiation field possesses only
monopole (from density or gravitational potential perturbations) and
dipole (from a Doppler shift) angular components; thus the radiation
is unpolarized. A quadrupole is subsequently produced at decoupling
as free streaming of the photons begins. A single Fourier mode of the
radiation field can be described by the temperature distribution
function $\Theta(k,\mu,\eta)$ where $k$ is the wavenumber, 
$\mu = {\bf\hat k}\cdot{\bf\hat n}$ is the angle between the 
vector $\bf k$ and the propagation direction $\bf\hat n$, 
and $\eta$ is conformal time. For mathematical simplicity only a flat
universe is considered here, although the non-flat cases are
no more complicated conceptually. Ignoring
gravitational potential contributions, free streaming of the photons
is described by the Liouville equation
$\dot\Theta + ik\mu\Theta = 0$. If the free streaming begins at
time $\eta_*$, then the solution at a later time is simply
$\Theta(k,\mu,\eta) = \Theta(k,\mu,\eta_*)\exp(-ik\mu(\eta - \eta_*))$.
Reexpressing the $\mu$ dependence as a multipole expansion,
\begin{equation}
\Theta(k,\mu,\eta) = \sum_{l=0}^\infty (-i)^l\Theta_l(k,\eta)P_l(\mu),
\end{equation}
the free streaming becomes
\begin{eqnarray}
\Theta_l(k,\eta) &=& (2l+1)[
\Theta_0(k,\eta_*)j_l(k\eta - k\eta_*)\nonumber\\
&&\qquad\qquad + \Theta_1(k,\eta_*)j'_l(k\eta - k\eta_*)],
\label{freestream}
\end{eqnarray}
where $j_l$ is the usual spherical
Bessel function.

We are interested in the behavior of the free streaming in the
at times near decoupling; at later times, the number density of
free electrons which can Thomson scatter has dropped to negligible
levels and no further polarization can be produced. The physical
length scales of interest for microwave background fluctuations
will be larger than the thickness of the last scattering surface,
so $k(\eta-\eta_*)$ will be small compared to unity. For small
arguments $x\ll 1$, $j_l(x)/j_l'(x)\sim x/l$, 
which implies that if the monopole
and dipole radiation components are initially of comparable size,
free streaming through the region of polarization generation 
with thickness $\Delta$ will generate
a quadrupole component from the dipole which is a factor of $2/(k\Delta)$
larger than the quadrupole component from the monopole. In other
words, on length scales large compared to the thickness of 
the surface of last scattering, the quadrupole moment and thus
the polarization couples much more strongly to the velocity
of the baryon-photon fluid than to the density.
Note that on smaller scales with $k\Delta\gtrsim 1$, the polarization
can couple more strongly to either the velocity or the density,
depending on the scale, but for standard recombination
these scales are always small enough
that the microwave background fluctuations are strongly
diffusion damped.

For the third part of the calculation, we need to know the mathematical
form for an acoustic oscillation. As emphasized in Ref.~\cite{husug},
the photon-baryon density perturbation in the tight-coupling
regime obeys the differential equation for a forced, damped
harmonic oscillator with the damping coming from the expansion
of the universe and the forcing from gravitational potential
perturbations. The solution is of the form
\begin{equation}
\Theta_0(k,\eta) =  A_1(\eta)\cos(kr_s) + A_2(\eta)\sin(kr_s)
\label{theta0}
\end{equation}
where the amplitudes vary slowly in time and $r_s\simeq \eta/\sqrt{3}$
is the sound horizon.
The velocity perturbation follows from the photon continuity
equation $\dot\Theta_0 = -k\Theta_1/3$, again neglecting gravitational
potential perturbations. A detailed consideration of boundary
conditions reveals that initial isentropic density perturbations couple
to the cosine harmonic in the small-scale limit, and this approximation
is good even for the largest-wavelength acoustic oscillations 
\cite{huwhite}.
Thus in an inflationary model, at the surface of last scattering,
the photon monopole has a $k$-dependence of approximately
$\cos(k\eta_*/\sqrt{3})$,
while the dipole, which is the main contributor to the polarization,
has a $k$-dependence of approximately $\sin(k\eta_*/\sqrt{3})$.
For initial isocurvature perturbations, the density perturbations
couple instead to the sine harmonic, but the photon monopole
and dipole are still $\pi/2$ out of phase.

From this scale dependence, it is simple to read off the relative
positions of the various acoustic peaks in the power spectra. In
general, the amplitude of the velocity perturbations are suppressed
by a factor of $c_s$ with respect to the density perturbations,
and the acoustic peaks in the power spectrum are dominated by
the monopole temperature fluctuation. The temperature 
power spectrum is roughly
the square of the temperature fluctuation, so acoustic peaks
in the power spectrum will show up at scales where 
$\cos^2(k\eta_*/\sqrt{3})$ has its maxima. Likewise, the
polarization couples to the temperature dipole on scales larger
than the thickness of the last scattering surface, and acoustic
peaks in the polarization power spectrum will be present at scales
where $\sin^2(k\eta_*/\sqrt{3})$ has its maxima, just the scales
between the temperature peaks. Finally, the cross-correlation
between the temperature and polarization will have extrema as
$-\cos(k\eta_*/\sqrt{3})\sin(k\eta_*/\sqrt{3})$ which fall
between the temperature and polarization peaks. (The correlation
between the polarization and the velocity contribution to the temperature
averages to zero because of their different angular dependences.)
All of these peak positions are evident in Fig.~1.
The sign of the cross-correlation peaks can be used to
deduce whether a temperature peak represents a compression or a
rarefaction, which can be checked against the alternating
peak-height signature
if the universe has a large enough baryon fraction \cite{husug}.

A combination of isentropic and isocurvature fluctuations shifts
all acoustic phases by the same amount if the ratio of their 
amplitudes is independent of scale, thus leaving the acoustic
signature intact. If the amplitude ratio depends on scale, the coherent
acoustic oscillations could be modified, but fine tuning would be 
required to wash them out completely. Multi-field inflation models
generically produce both isocurvature and isentropic perturbations
\cite{steinhardt} but the resulting microwave background power spectra
are just beginning to be studied in detail \cite{kanazawa}.

Inflation unavoidably produces coherent acoustic oscillations
due to isentropic perturbations on all scales. 
Defect models can in principle generate coherent oscillations, but
the evolution of the defects must not destroy the phase coherence
of a given oscillation mode \cite{decoherence} so the resulting
models are rather artificial. One example is the large-N limit of 
the O(N) $\sigma$-model \cite{sigma}. But despite freedom to choose five
functions defining a general causal scaling solution for the evolution
of a defect \cite{durrer}, it seems highly unlikely that an 
inflation-like set of $C_l$ spectra can be closely matched
\cite{nodefects}.
Polarization, coupling only to velocities and not to density
or gravitational fluctuations, provides a particularly stringent
discriminator. It seems virtually impossible for any model with
causally generated sources to reproduce the small first polarization
acoustic peak (around $l\simeq 80$ in a flat universe) \cite{spergel}.

So polarization provides a model-independent signature of coherent
acoustic oscillations at the time of last scattering: acoustic
peaks in the polarization power spectrum should fall almost precisely
between the acoustic peaks in the temperature power spectrum, and
the extrema of the temperature-polarization cross-correlation 
fall between the temperature and polarization acoustic peaks. This
signature is the easiest polarization information to extract from 
the microwave background since it depends only on the 
largest-amplitude features of the polarization fluctuations.
Hopefully the prospect of seeing sound waves in the early universe 
will provide
further impetus to ongoing efforts to detect microwave background
polarization.

\acknowledgments
I thank Ruth Durrer and Suzanne Staggs for helpful conversations,
and Peter Timbie for comments on a draft version.
This work has been supported by NASA Theory Program
through grant NAG5-7015. Portions of this work were
done at the Institute for Advanced Study.


\begin{references}

\bibitem{parameters} See, e.g., G.~Jungman et al., Phys.\ Rev.\ D
     54, 1332 (1996);
     M.~Zaldarriaga, D.N.~Spergel, and U.~Seljak, Astrophys.\ J.
     488, 1 (1997); J.R.~Bond, G.~Efstathiou, and M.~Tegmark, 
     Mon.\ Not.\ R.\ Ast.\ Soc.\ 291, L33 (1997); 
     J.R.~Bond and G.~Efstathiou, astro-ph/9807103;
     D.J.~Eisenstein, W.~Hu, and M.~Tegmark, astro-ph/9807130.

\bibitem{cmbfast} Power spectra were generated with the
     CMBFAST code, U.~Seljak and M.~Zaldarriaga, Astrophys.\ J.
     469, 437 (1996); 
     http://www.sns.ias.edu/\~{}matiasz/CMBFAST/cmbfast.html.

\bibitem{annals} For a similar discussion, see A.~Kosowsky,
     Ann.\ Phys.\ 246, 49 (1996).


\bibitem{primer} W.~Hu and M.~White, New Astron.\ 2, 323 (1997).

\bibitem{husug} W.~Hu and N.~Sugiyama, Astrophys.\ J.\ 444, 489 (1995).

\bibitem{huwhite} W.~Hu and M.~White, Astrophys.\ J.\ 471, 30 (1996).

\bibitem{steinhardt} V.F.~Mukhanov and P.J.~Steinhardt, Phys.\ Lett.\
     B422, 52 (1998). 

\bibitem{kanazawa} T.~Kanazawa et al., astro-ph/9805102.

\bibitem{decoherence} For power spectrum computations in models
     with substantial decoherence, see R.~Durrer and M.~Sakellariadou,
     Phys.\ Rev.\ D 56, 4480 (1997); U.~Pen, U.~Seljak and
     N.~Turok, Phys.\ Rev.\ Lett.\ 79, 1611 (1997); U.~Seljak,
     U.~Pen, and N.~Turok, Phys.\ Rev.\ Lett.\ 79, 1615 (1997); 
     J.~Magueijo et al., Phys.\ Rev.\ D 54, 2737 (1996);
     B.~Allen et al., Phys.\ Rev.\ Lett.\ 79, 2624 (1997);
     C.~Contaldi, M.~Hindmarsh, and J.~Magueijo,
     astro-ph/9808201.

\bibitem{sigma} N.~Turok and D.N.~Spergel, 
     Phys.\ Rev.\ Lett.\ 66, 3093 (1991);
     A.~Jaffe, Phys.\ Rev.\ D 49, 3893 (1994); 
     M.~Kunz and R.~Durrer, Phys.\ Rev.\ D 55, R4516 (1997).

\bibitem{durrer} R.~Durrer and M.~Kunz, Phys.\ Rev.\ D 57, R3199 (1998).

\bibitem{nodefects} W.~Hu, D.N.~Spergel, and M.~White, Phys.\ Rev.\ D
     55, 3288 (1997).

\bibitem{spergel} M.~Zaldarriaga and D.N.~Spergel, Phys.\ Rev.\
     Lett.\ 79, 2180 (1997).

\end{references}
\end{document}